\documentclass[article]{JHEP3} 

\usepackage{epsfig,multicol,bbm}
\newcommand\fverb{\setbox\fverbbox=\hbox\bgroup\verb}
\newcommand{\LL}{Lanczos-Lovelock }
\newcommand\fverbdo{\egroup\medskip\noindent%
			\fbox{\unhbox\fverbbox}\ }
\newcommand\fverbit{\egroup\item[\fbox{\unhbox\fverbbox}]}
\newbox\fverbbox

\title{Noether current of the surface term of Einstein-Hilbert action, Virasoro algebra and entropy}

\author{Bibhas Ranjan Majhi\\
IUCAA, Post Bag 4, Ganeshkhind,
Pune University Campus, Pune 411 007, India \\
	E-mail: \email{bibhas@iucaa.ernet.in}}

\abstract{A derivation of Noether current from the surface term of Einstein-Hilbert action is given. We show that the corresponding charge, calculated on the horizon, is related to the Bekenstein-Hawking entropy. Also using the charge, the same entropy is found based on the Virasoro algebra and Cardy formula approach. In this approach, the relevant diffeomorphisms are found by imposing a very simple physical argument: {\it diffeomorphisms keep the horizon structure invariant}. This complements similar earlier results \cite{Majhi:2012tf}(arXiv:1204.1422) obtained from York-Gibbons-Hawking surface term. Finally we discuss the technical simplicities and improvements over the earlier attempts and also various important physical implications.}


\begin{document} 
\section{Introduction}
   The thermodynamic properties of horizon arises from the combination of the general theory of relativity and the quantum field theory. This was first observed in the case of black holes \cite{Bekenstein:1973ur,Hawking:1974rv}. Now it is evident that it is much more general and a local Rindler observer can attribute temperature and entropy to the null surfaces in the context of the emergent paradigm of gravity \cite{TPreviews,Banerjee:2010yd}. Such a generality might provide us a deeper insight towards the quantum nature of the spacetime. So far several attempts have been made to know the microscopic origin of the entropy, but every method has its own merits and demerits. Among others, Carlip made an attempt \cite{Carlip:1998wz,Carlip:1999cy} in the context of Virasoro algebra to illuminate this aspect which is basically the generalisation of the method by Brown and Hanneaux \cite{Brown:1986nw}. In brief, in this method one first defines a bracket among the Noether charges and calculate it for certain diffeomorphisms, chosen by some physical considerations. It turns out that the algebra is identical to the Virasoro algebra. The central charge and the zero mode eigenvalue of the Fourier modes of the charge are then automatically identified which after substituting in the Cardy formula \cite{Cardy:1986ie,Carlip:1998qw} one finds the Bekenstein-Hawking entropy {\footnote{For a complete list of works which lead to further development of this method, see \cite{Solodukhin:1998tc,Brustein:2000fw,Majhi:2012st,Lin:1999gf,Das:2000zs,Jing:2000yn,Terashima:2001gn,Koga:2001vq,Park:2001zn,Carlip:2002be,Cvitan:2002cs,Cvitan:2002rh,Park:1999tj,Park:1999hs,Dreyer:2001py,Silva:2002jq}.}}.  In all the previous attempts, the Noether current was taken related to the Einstein-Hilbert (EH) action and the analysis was on-shell, i.e. equation of motion has been used explicitly. Later an off-shell analysis and a generalization to \LL gravity have been presented in \cite{Majhi:2011ws}.
\vskip 2mm
  Earlier \cite{Majhi:2012tf}, based on the Virasoro algebra approach, we showed that the entropy can also be obtained from the Noether current corresponding to the York-Gibbons-Hawking surface term. But it is not clear if the same can be achived from the surface term of the Einstein-Hilbert action, since they are not exactly identical. So it is necessary to investigate this issue in the light of Virasoro algebra context, particulary because both the surface terms lead to the same entropy on the horizon. This will complement our earlier work \cite{Majhi:2012tf}.

\vskip 2mm
  In this paper we will use the Noether current associated to the surface term of EH action. Before going into the motivations for taking the surface term only, let us first highlight some peculiar facts of EH action which are essential for the present purpose. 
\vskip 0.1mm
\noindent
$\bullet$ It is an unavoidable fact that to obtain the equation of motion in the Lagrangian formalism one has to impose some extra prescription, like adding extra boundary term (in this case York-Gibbons-Hawking term). This is because the action contains second order derivative of metric tensor $g_{ab}$. But unfortunately the choice of the surface term is not unique. This is quite different from other well known field theories.
\vskip 0.1mm
\noindent
$\bullet$ The EH action can be separated into two terms: one contains the squires of the Christoffel connections ( i. e. it is in $\Gamma\Gamma - \Gamma\Gamma$ structure) and the other one contains the total derivation of $\Gamma$ ($\partial \Gamma-\partial\Gamma$ structure). We will call them as $L_{quad}$ and $L_{sur}$, respectively. Interestingly, Einstein's equation of motion can be obtained solely from $L_{quad}$ by using the usual variation principle where no additional prescription is not required \cite{Padmanabhan:2010zzb}.
\vskip 0.1mm
\noindent
$\bullet$ The most important one is that these two terms are related by an algebraic relation, usually known as {\it holographic} relation \cite{Padmanabhan:2004fq,Mukhopadhyay:2006vu}.
\vskip 0.1mm
\noindent
Interestingly, all the above features are happened to be common even for the Lanczos-Lovelock theory \cite{Kolekar:2010dm}. For a recent review in this direction, see \cite{Padmanabhan:2012qz}.
\vskip 2mm

  Although an extensive study on the Noether current of gravity has been done starting from Wald \cite{Wald:1993nt}, discussion on the current derived from $L_{sur}$ is still lacking. To motivate why one should be interested, let us summarise below the already observed facts.  
\vskip 0.1mm
\noindent
$\bullet$ It is expected that the entropy is associated to the degrees of freedom around or on the relevant null surface rather than the bulk geometry of spacetime.
\vskip 0.1mm
\noindent
$\bullet$ This surface term calculated on the Rindler horizon gives exactly the Bekenstein-Hawking entropy \cite{Padmanabhan:2010zzb}.
\vskip 0.1mm
\noindent
$\bullet$ Extremization of the surface term with respect to the diffeomorphism parameter whose norm is a constant, leads to the Einstein's equation \cite{Padmanabhan:2004fq}.
\vskip 0.1mm
\noindent
$\bullet$ Another interesting fact is that in a small region around an event, EH action reduces to a pure surface term when evaluated in the Riemann normal coordinates.
\vskip 0.1mm
\noindent
All these indicate that either the bulk and the surface terms are duplicating all the information or the actual dynamics is stored in surface term rather than in bulk term. To illuminate more on this issue, one needs to study every aspect of the surface term. 

\vskip 2mm
   In this paper, we shall discuss the Noether realization of the surface term of the EH action, particularly we shall examine if the Noether current represents the Virasoro algebra for a certain class of diffeomorphisms. This is necessary to have a deeper understanding of the role of the surface term in the gravity. Also it will give a further insight towards the earlier claim: {\it the actual information of the gravity is stored in the surface}. To do this explicitly, we shall consider the form of the metric close to the null surface in the local Rindler frame around some event. This is given by the Rindler metric. The reasons for choosing such metric are as follows. According to equivalence principle, gravity can be mimicked by an accelerated observer and an uniformly accelerated frame will have Rindler metric. Apart from that, it is a relevant frame for an observer sitting very near to the black hole horizon. Hence any thermodynamic feature of the null surface can be attributed by this metric and it provides a general description which was originally obtained only for the black hole horizon. Moreover, all the quantities will be observer dependent.

\vskip 2mm      
   In this paper we shall proceed as follows. First a detailed derivation of the Noether current for a diffeomorphism $x^a\rightarrow x^a+\xi^a$, corresponding to $L_{sur}$, will be given. This is important because it has not been done earlier and therefore the properties of the current have not been explored. Here we will show that the corresponding charge $Q[\xi]$, calculated on the null surface for $\xi^a$ to be Killing, yields exactly one quarter of the horizon area after multiplying it by $2\pi/\kappa$ where $\kappa$ is the acceleration of the observer or the surface gravity in the case of a black hole.  Next, a definition of the bracket among the charges will be given. This will be done by taking variation of the charge $Q[\xi_1]$ for another transformation $x^a\rightarrow x^a+\xi_2^a$. Finally, we need to calculate all these quantities for a particular diffeomorphism. To identify the relevant diffeomorphisms from which the algebra has to be constructed, following our earlier work \cite{Majhi:2012tf}, we use the criterion that the diffeomorphism should leave the near horizon form of the metric invariant in
some non-singular coordinate system. This will lead to a set of diffeomorphism vectors for which the Fourier components of the bracket among the charges will be exactly similar to Virasoro algebra. It is then very easy to identify the zero mode eigenvalue and the central extension. Substitution of all these values in the Cardy formula \cite{Cardy:1986ie,Carlip:1998qw} will yield exactly the Bekenstein-Hawking entropy \cite{Bekenstein:1973ur,Hawking:1974rv}. A similar analysis was done in \cite{Silva:2002jq} based on the Noether current corresponding to $L_{bulk}$ \cite{Julia:1998ys,Julia:2000er,Silva:2000ys}. In this calculation, to obtain the correct value of the entropy, a particular boundary condition (Dirichlet or Neumann) was used. But the physical significance of it is not well understood.

\vskip 2mm
  Before going into the main calculation, let us summarize the main features of the present analysis.
\vskip 0.1mm
\noindent
$\bullet$ First is the technical aspect. To obtain the correct entropy, in most of the earlier works, one had to either shift the zero mode eigenvalue \cite{Carlip:1999cy} or choose a parameter contained in the Fourier modes of $\xi^a$ as the surface gravity $\kappa$ \cite{Silva:2002jq} or both \cite{Majhi:2011ws}. Here we shall show that none of ad hoc prescriptions will be required.
\vskip 0.1mm
\noindent
$\bullet$ The important one is the simplicity of the criterion ({\it near horizon structure of the metric remains invariant in some non-singular coordinate system}) to find the relevant diffeoprphisms for which we obtain the Virasoro algebra. This was first introduced by us \cite{Majhi:2012tf} in this context. The significance of this choice is that the full set of diffeomorphism symmetry of the theory is now reduced to a subset which respects the existence of horizon in a given coordinate system. Hence it may happen that some of the original gauge degrees of freedom (which could have been eliminated by certain diffeomorphisms which are now disallowed) now being effectively upgraded to physical degrees of freedom as far as a particular class of observers are concerned. So all the thermodynamic quantities, attributed to the horizon, become observer dependent.
\vskip 0.1mm
\noindent
$\bullet$ In our present analysis we will not need any use of boundary condition like Dirichlet or Neumann to obtain the exact form of the entropy.  
\vskip 0.1mm
\noindent
$\bullet$ Since our analysis will be completely based on $L_{sur}$ where no information about $L_{bulk}$ is needed, it will definitely illuminate the emergent paradigm of gravity, particularly the holographic aspects in the action.
\vskip 0.1mm
\noindent 
We will discuss later more on different aspects and significance of our results.

\vskip 2mm
The organization of the paper as follows. In section 2, the derivation of the Noether current for the $L_{sur}$ will be presented explicitly. Next we shall give the definition of the bracket among the charges and the relevant diffeomorphims based on the invariance of horizon structure criterion. Section 4 will be devoted to show that the Fourier mode of the bracket is exactly like the Virasoro algebra which by the Cardy formula will lead to Bekenstein-Hawking entropy. Finally, we shall conclude. 
 
\section{Derivation of Noether current from the surface term of Einstein-Hilbert action}
   In this section, a detailed derivation of the Noether current and the potential corresponding to the surface term of EH action will be presented. Then we shall calculate the charge on the Rindler horizon.

   The Lagrangian corresponding to the surface term is given by \cite{Padmanabhan:2010zzb},
\begin{eqnarray}
L_{sur} = \partial_a(\sqrt{-g}S^a)~,
\label{App1}
\end{eqnarray}
where
\begin{eqnarray}
S^a = 2Q^{ad}_{ck}g^{bk}\Gamma^c_{bd}; \,\,\,\ Q^{ad}_{ck} = \frac{1}{2} (\delta^a_c\delta^d_k - \delta^a_k\delta^d_c)~.
\label{App1.01}
\end{eqnarray}
Here the normalization $1/16\pi G$ is omitted and it will be inserted where necessary.
Now our task is to find the variations of both sides of (\ref{App1}) for a diffeomorphism $x'^a = x^a+\xi^a$ and then equate them. The variation we shall consider here as the Lie variation which is defined, in general, as
\begin{eqnarray}
\delta A = A(x') - A'(x')~,
\label{App1.02}
\end{eqnarray}
where $A(x') = A(x+\xi) = A(x) + \xi^a\partial_a A(x)$, $A(x)$ and $A'(x')$ are the evaluated in two different coordinate systems $x$ and $x'$, respectively. In the following, for the notational simplicity, we shall denote $A(x)$ as $A$.

  The variation of the right hand side of (\ref{App1}) is given by,
\begin{eqnarray}
\delta L_{sur} &=& \partial_a[\delta(\sqrt{-g}S^a)] =  \partial_a\Big[ S^a\delta(\sqrt{-g})+\sqrt{-g}\delta S^a\Big]
\nonumber
\\
&=& \partial_a\Big[\frac{S^a}{2}\sqrt{-g}g^{bc}\delta g_{bc} + \sqrt{-g}\delta S^a\Big]~.
\label{App1.03}
\end{eqnarray}
Since $g_{ab}$ is a tensor, for the Lie variation, $\delta g_{ab}$ is expressed by the Lie derivative and is given by 
\begin{eqnarray}
\delta g_{ab} = \nabla_a\xi_b + \nabla_b\xi_a~.
\label{App1.04}
\end{eqnarray}
Therefore, 
\begin{eqnarray}
\delta L_{sur} &=& \partial_a\Big[ S^a \partial_b(\sqrt{-g}\xi^b)+\sqrt{-g}\delta S^a\Big]~.
\label{App5}
\end{eqnarray}
On the other hand, since $S^a$ is not a tensor, the variation of it can not be expressed by simple Lie derivative. To find $\delta S^a$ we shall used the general definition (\ref{App1.02}). Let us first calculate $S'^a(x')$. Under the change $x'^a=x^a+\xi^a$ we have,
\begin{eqnarray}
&&\frac{\partial x'^a}{\partial x^b} = \delta^a_b + \partial_b \xi^a~;
\nonumber
\\
&& \frac{\partial x^b}{\partial x'^a} = \delta^b_a - \partial_a \xi^b~.
\label{App6}
\end{eqnarray}
Here we considered infinitesimal change and so the terms from $\partial\xi\partial\xi$ have been ignored. This will be followed in the later analysis.
Hence,
\begin{eqnarray}
&&\Gamma'^a_{bc}(x') = \Gamma^a_{bc} - \Gamma^a_{bd}\partial_c\xi^d - \Gamma^a_{cd}\partial_b\xi^d + \Gamma^d_{bc}\partial_d\xi^a - \partial_b\partial_c\xi^a~,
\nonumber
\\
&&g'^{bk}(x') = g^{bk} + g^{bf}\partial_f\xi^k + g^{kf}\partial_f\xi^b~,
\nonumber
\\
&&Q'^{ad}_{ck} (x')= Q^{ad}_{ck}~.
\label{App7}
\end{eqnarray}
Substitution of these in $S'^a(x') = 2Q'^{ad}_{ck}(x')g'^{bk}(x')\Gamma'^c_{bd}(x')$ lead to,
\begin{eqnarray}
S'^a(x')=  S^a + S^b\partial_b\xi^a - g^{bd}\partial_b\partial_d\xi^a + g^{ab}\partial_b\partial_c\xi^c~.
\label{App8}
\end{eqnarray}
Other one is given by
\begin{eqnarray}
S^a(x') = S^a(x^b+\xi^b) =  S^a + \xi^b\partial_b S^a~.
\label{App9}
\end{eqnarray}
Therefore, according to (\ref{App1.02}), the Lie variation of $S^a$ due to the diffeomorphism is
\begin{eqnarray}
\delta S^a = S^a(x') - S'^a(x') = \xi^b\partial_b S^a - S^b\partial_b\xi^a + M^a~,
\label{App10}
\end{eqnarray}
where
\begin{eqnarray}
M^a = g^{bd}\partial_b\partial_d\xi^a - g^{ab}\partial_b\partial_c\xi^c~.
\label{App11}
\end{eqnarray}
Substituting this in (\ref{App5}) we obtain the variation of right hand side of (\ref{App1}) as,
\begin{eqnarray}
\delta L_{sur} = \partial_a\Big[\partial_b(\sqrt{-g}S^a\xi^b) - \sqrt{-g}S^b\partial_b\xi^a + \sqrt{-g}M^a\Big]~.
\label{App12}
\end{eqnarray}

  Next we find the variation of left hand side of (\ref{App1}); i.e. $L_{sur}$. For this we will start from the following relation:
\begin{eqnarray}
L_{sur} = \sqrt{-g}(L_g - L_{quad})~,
\label{App13}
\end{eqnarray}
where 
\begin{eqnarray}
L_g = R; \,\,\,\ L_{quad} = 2Q^{bcd}_a\Gamma^a_{dk}\Gamma^k_{bc}~,
\label{App14}
\end{eqnarray}
with $Q^{bcd}_a= \frac{1}{2}\Big(\delta^c_a g^{bd} - \delta^d_a g^{bc}\Big)$.
Since $L_g$ is a scalar, by the definition of Lie derivative $\delta L_g = \xi^a\partial_a L_g$. Therefore using (\ref{App1.04}) we find
\begin{eqnarray}
\delta L_{sur} &=& \delta(\sqrt{-g}L_g) - \delta(\sqrt{-g}L_{quad})
\nonumber
\\
&=& \partial_a(\sqrt{-g}\xi^aL_g) - \partial_a(\sqrt{-g}\xi^a)L_{quad} - \sqrt{-g}\delta L_{quad}
\nonumber
\\
&=& \partial_a\Big[\sqrt{-g}\xi^a(L_g - L_{quad})\Big] + \sqrt{-g}\xi^a\partial_aL_{quad} - \sqrt{-g}\delta L_{quad}
\nonumber
\\
&=& \partial_a\Big(\xi^a L_{sur}\Big) + \sqrt{-g}\xi^a\partial_aL_{quad} - \sqrt{-g}\delta L_{quad}~.
\label{App15}
\end{eqnarray}
To find $\delta L_{quad}$, we will proceed as earlier. Under the change $x'^a = x^a+\xi^a$, $L'_{quad}(x')$ is calculated as
\begin{eqnarray}
L'_{quad} (x') &=& 2 Q'^{bcd}_a(x') \Gamma'^a_{dk}(x')\Gamma'^k_{bc}(x')
\nonumber
\\
&=& L_{quad} + g^{bc}\Gamma^k_{bc}\partial_d\partial_k\xi^d + g^{bc}\Gamma^d_{dk}\partial_b\partial_c\xi^k - g^{bd}\Gamma^c_{dk}\partial_b\partial_c\xi^k~,
\label{App16}
\end{eqnarray}
where (\ref{App7}) has been used. This can be expressed in terms of $M^a$ in the following way.
Second term on the right hand side can be expressed in the following form
\begin{eqnarray}
\sqrt{-g}g^{bc}\Gamma^k_{bc}\partial_d\partial_k\xi^d &=& \Big[\sqrt{-g}g^{bc}g^{ak}\partial_bg_{ac} - g^{ak}\partial_a(\sqrt{-g})\Big]\partial_d\partial_k\xi^d
\nonumber
\\
&=& - \partial_a(\sqrt{-g}g^{ak})\partial_d\partial_k\xi^d~,
\label{App18}
\end{eqnarray}
where in the above we used $g^{bc}g^{ak}\partial_bg_{ac} = -\partial_ag^{ak}$. Third term of (\ref{App16}) reduces to
\begin{eqnarray}
\sqrt{-g} g^{bc}\Gamma^d_{dk}\partial_b\partial_c\xi^k = \partial_k(\sqrt{-g})g^{bc}\partial_b\partial_c\xi^k~.
\label{App19}
\end{eqnarray}
Similarly, the last term can be expressed as
\begin{eqnarray}
2 \sqrt{-g}g^{bd}\Gamma^c_{dk}\partial_b\partial_c\xi^k &=& \sqrt{-g}g^{bd}g^{ca}\partial_kg_{ad}\partial_b\partial_c\xi^k
\nonumber
\\
&=& - \sqrt{-g}\partial_k(g^{bc})\partial_b\partial_c\xi^k~,
\label{App20}
\end{eqnarray}
where in the last line $g^{bd}g^{ca}\partial_kg_{ad} = - \partial_kg^{bc}$ has been used. Substituting all these in (\ref{App16}) we obtain
\begin{eqnarray}
L'_{quad}(x') &=& L_{quad} - \frac{1}{\sqrt{-g}} \Big[\partial_a(\sqrt{-g}g^{ak})\partial_d\partial_k\xi^d + \partial_k(\sqrt{-g}g^{bc})\partial_b\partial_c\xi^k\Big]
\nonumber
\\
&=& L_{quad} + \frac{1}{\sqrt{-g}} \partial_a(\sqrt{-g}M^a)
\label{App21}
\end{eqnarray}
On the other hand,
\begin{eqnarray}
L_{quad} (x')= L_{quad}(x^a+\xi^a) =   L_{quad} + \xi^a\partial_a L_{quad}~.
\label{App22}
\end{eqnarray} 
Hence
\begin{eqnarray}
\sqrt{-g}\delta L_{quad} = \sqrt{-g} L_{quad}(x') - \sqrt{-g}L'_{quad}(x') = \sqrt{-g}\xi^a\partial_a L_{quad} - \partial_a(\sqrt{-g}M^a)~.
\label{App23}
\end{eqnarray}
Substituting this in (\ref{App15}) we obtain
\begin{eqnarray}
\delta L_{sur} = \partial_a\Big(\xi^aL_{sur} + \sqrt{-g}M^a\Big)~.
\label{App24}
\end{eqnarray}
Now equating (\ref{App12}) and (\ref{App24}) we obtain $\partial_a J^a[\xi] = 0$, where the conserved Noether current $J^a[\xi]$ is given by
\begin{eqnarray}
J^a[\xi] = -\partial_b(\sqrt{-g}S^a\xi^b) + \sqrt{-g}S^b\partial_b\xi^a  + \xi^a L_{sur}~.
\label{App25}
\end{eqnarray}
Finally, using $L_{sur} = \partial_a(\sqrt{-g}S^a)$ in the above, we can express the current as the divergence of a anti-symmetric two index quantity:
\begin{eqnarray}
J^a[\xi] = \partial_b\Big[\sqrt{-g}(\xi^aS^b - \xi^bS^a)\Big]=\partial_b\Big[\sqrt{-g}J^{ab}[\xi]\Big]~.
\label{App26}
\end{eqnarray}
It is evident that the anti-symmetric object $J^{ab}[\xi]$ is not a tensor and it is usually called the Noether potential.
Therefore, inserting the proper normalization, the charge is given by
\begin{eqnarray}
Q[\xi] = \frac{1}{32\pi G} \int_{\cal{H}} d\Sigma_{ab} \sqrt{h} J^{ab}[\xi]~,
\label{1.47}
\end{eqnarray}
where $d\Sigma_{ab} = - d^2x(N_a M_b - N_b M_a)$ is the surface
element of the $2$-dimensional surface $\cal{H}$ and $h$ is
the determinant of the corresponding metric. Since our
present discussion will be near the horizon, we choose
the unit normals $N_a$ and $M_a$ as spacelike and timelike
respectively.

Now we shall calculate the charge (\ref{1.47}) explicitly on the horizon. This will be done by considering the form of the metric near the horizon, 
\begin{eqnarray}
ds^2 = -2\kappa x dt^2 + \frac{1}{2\kappa x} dx^2 + dx^2_{\perp}~,
\label{1.03}
\end{eqnarray}
where $x_\perp$ represents the transverse coordinates.
The metric has a timelike Killing vector $\chi^a = (1,0,0,0)$ and the Killing horizon is given by $\chi^2 = 0$; i.e. $x = 0$.
The non-zero Christoffer connections are
\begin{eqnarray}
\Gamma^t_{tx} = \frac{1}{2x};\,\,\,\ \Gamma^x_{tt} = 2\kappa^2 x ;\,\,\,\ \Gamma^x_{xx}=-\frac{1}{2x}~.
\label{1.06}
\end{eqnarray}
For the metric (\ref{1.03}) we find
\begin{eqnarray}
N^a = (0,\sqrt{2\kappa x},0,0); \,\,\ M^a = (\frac{1}{\sqrt{2\kappa x}},0,0,0)~,
\label{App27}
\end{eqnarray}
and hence $d\Sigma_{tx} = - d^2x$. Also, (\ref{App1.01}) yields
\begin{eqnarray}
S^t = 0; \,\,\ S^x = -2\kappa~.
\label{1.51}
\end{eqnarray}
Therefore, 
\begin{eqnarray}
J^{tx} =  ( \xi^t  S^x - \xi^x  S^t) = -2\kappa\xi^t~.
\label{1.50}
\end{eqnarray}
Now if $\xi^a$ is a Killing vector, then $\xi^t = \chi^t = 1$ and so calculating the charge (\ref{1.47}) explicitly we find
\begin{eqnarray}
Q[\xi=\chi] = \frac{\kappa A_{\perp}}{8\pi G}~,
\label{App28}
\end{eqnarray}
where $A_{\perp} = \int_{\cal{H}}d^2x$ is the horizon cross-section area. Multiplying it by the periodicity of time coordinate $2\pi/\kappa$ we obtain exactly the entropy: one quarter of horizon area. Moreover, the above can be expressed as $Q[\xi=\chi]=TS$, where $T=\kappa/2\pi$ is the temperature of the horizon and $S = A_{\perp/4 G}$ is the entropy. Therefore one can call it as the Noether energy. Such interpretaion was done earlier in \cite{Ashworth:1998uj,Padmanabhan:2012bs}.  

   So far we found that the Noether charge corresponding to the surface term of EH action alone led to the entropy of the Rindler horizon. This was shown earlier for the charge coming from the total EH action \cite{Wald:1993nt}. Therefore, the present analysis reveled that it may be possible that the information is actually encoded in the surface term rather than the bulk term. Then the natural question arises: What are the degrees of freedom responsible for this entropy? So far it is not known. In the next couple of sections we shall give an idea on the nature of the possible degrees of freedom in the context of Virasoro algebra and Cardy formula.

\section{Bracket among the charges and the diffeomorphism generators}
   In the previous section, we have given the expression for the charge (see Eq. (\ref{1.47})) for an arbitrary diffeomorphism. Here we shall define the bracket among the charges. The relevant diffeomorphisms will be chosen by imposing a minimum condition on the spacetime metric. The charge and the bracket will be then expressed in terms of these generators.

  We shall find the bracket following our earlier works \cite{Majhi:2011ws,Majhi:2012tf}. For this let us first calculate the following:
\begin{eqnarray}
\delta_{\xi_1}(\sqrt{-g}J^{ab}[\xi_2]) &=& \delta_{\xi_1}(\sqrt{-g})J^{ab}[\xi_2] + \sqrt{-g} \delta_{\xi_1}(J^{ab}[\xi_2])
\nonumber
\\
&=&-\frac{1}{2}\sqrt{-g} g_{mn}\delta_{\xi_1}g^{mn} J^{ab}[\xi_2] 
\nonumber
\\
&+& \sqrt{-g}\Big[(\delta_{\xi_1}\xi_2^a) S^b + \xi_2^a (\delta_{\xi_1}S^b) - (a\leftrightarrow b)\Big]~.
\label{1.53}
\end{eqnarray}
Using
\begin{eqnarray}
\delta_{\xi}g^{ab} &=& \pounds_{\xi}g^{ab} = -\nabla^a\xi^b - \nabla^b\xi^a;
\nonumber
\\
\delta_{\xi}\Gamma^a_{bc} &=& \nabla_b\nabla_c\xi^a + R^a_{cmb}\xi^m~,
\label{1.54}
\end{eqnarray}
and in addition the expression for $S^a$, given by (\ref{App1.01}), we obtain,
\begin{eqnarray}
\delta_{\xi_1}(\sqrt{-g}J^{ab}[\xi_2]) &=& \sqrt{-g} \Big[\nabla_m\xi_1^m J^{ab}[\xi_2] + \{(\xi_1^m\nabla_m\xi_2^a - \xi_2^m\nabla_m\xi_1^a)S^b
\nonumber
\\
&+& \xi_2^a\Big(-2\Gamma^b_{mn}\nabla^m\xi_1^n + \nabla_m\nabla^m\xi_1^b + 2R^b_m\xi_1^m 
\nonumber
\\
&-& \Gamma^n_{nm}(\nabla^b\xi_1^m + \nabla^m\xi_1^b) - \nabla_m\nabla^b\xi_1^m \Big)- (a\leftrightarrow b)\}\Big]~.
\label{1.55}
\end{eqnarray}
For the present metric (\ref{1.03}), $g = -1$, $R^a_b = 0$ and hence $\sqrt{-g}\Gamma^n_{nm} = \partial_m(\sqrt{-}g) = 0$. Therefore
\begin{eqnarray}
\delta_{\xi_1}(\sqrt{-g}J^{ab}[\xi_2]) &=& \Big[\nabla_m\xi_1^m J^{ab}[\xi_2] + \frac{1}{16\pi G}\{(\xi_1^m\nabla_m\xi_2^a - \xi_2^m\nabla_m\xi_1^a)S^b
\nonumber
\\
&+& \xi_2^a\Big(-2\Gamma^b_{mn}\nabla^m\xi_1^n + \nabla_m\nabla^m\xi_1^b
 - \nabla_m\nabla^b\xi_1^m \Big) - (a\leftrightarrow b)\}\Big]
\nonumber
\\
\equiv K^{ab}_{12} ~.
\label{1.56}
\end{eqnarray}
Finally we define a bracket as:
\begin{eqnarray}
[Q[\xi_1],Q[\xi_2]]: = \frac{1}{2} \int_{\cal{H}} d\Sigma_{ab}\sqrt{h}\Big[K^{ab}_{12} - (1\leftrightarrow 2)\Big]~,
\label{1.48}
\end{eqnarray}
which for the present metric (\ref{1.03}) reduces to
\begin{eqnarray}
[Q[\xi_1],Q[\xi_2]]: =  - \int_{\cal{H}} d^2x \Big[K^{tx}_{12} - (1\leftrightarrow 2) \Big]~.
\label{1.57}
\end{eqnarray}

   To calculate the above bracket we need to know about the generators $\xi^a$. We shall determine them by using the condition that the horizon structure remains invariant in some nonsingular coordinate system.  For that let us first express the metric (\ref{1.03}) in Gaussian null (or Bondi like) coordinates,
\begin{eqnarray}
du = dt - \frac{dx}{2\kappa x}; \,\,\ dX &=& dx~.
\label{1.07}
\end{eqnarray}
In these coordinates the metric reduces to the following form:
\begin{eqnarray}
ds^2 = -2\kappa X du^2 - 2 du dX + dx^2_{\perp}~.
\label{1.08}
\end{eqnarray}
Now impose the condition that the metric coefficients $g_{XX}$ and $g_{uX}$ do not change under the diffeomorphism, i.e.
\begin{eqnarray}
\pounds_{\tilde{\xi}} g_{XX} = 0; \,\,\,\ \pounds_{\tilde{\xi}} g_{uX} =0~,
\label{1.11}
\end{eqnarray}
where $\pounds_{\tilde{\xi}}$ is the Lie derivative along the vector $\tilde{\xi}$. These lead to,
\begin{eqnarray}
&&\pounds_{\tilde{\xi}} g_{XX} = -2\partial_X\tilde{\xi}^u = 0;
\nonumber
\\
&& \pounds_{\tilde{\xi}} g_{uX} = -\partial_u \tilde{\xi}^u - 2\kappa X \partial_X \tilde{\xi}^u - \partial_X \tilde{\xi}^X =0~.
\label{1.12}
\end{eqnarray}
The solutions are:
\begin{eqnarray}
&&\tilde{\xi}^u = F(u, x_{\perp});
\nonumber
\\
&& \tilde{\xi}^X = - X\partial_u F(u,x_{\perp})~.
\label{1.13}
\end{eqnarray}
The condition $\pounds_{\tilde{\xi}} g_{uu} = 0$ automatically satisfied near the horizon, because use of the above solutions lead to $\pounds_{\tilde{\xi}} g_{uu} = {\cal{O}}(X)$. These conditions were appeared earlier in \cite{Tanabe:2011zt} in the context of late time symmetry near the black hole horizon.
Finally expressing (\ref{1.13}) in the old coordinates ($t,x$) we find
\begin{eqnarray}
\xi^t = T - \frac{1}{2\kappa}\partial_tT; \,\,\,\ \xi^x = -x\partial_t T~,
\label{1.17}
\end{eqnarray}
where $T(t,x,x_{\perp}) = F(u,x_{\perp})$.

  Next we calculate $K^{tx}_{12}$ from (\ref{1.56}) for our present case.   
Since $S^t =0$, we find
\begin{eqnarray}
 K^{tx}_{12} &=& \nabla_m\xi_1^m \xi_2^t S^x +(\xi_1^m\nabla_m\xi_2^t - \xi_2^m\nabla_m\xi_1^t)S^x
\nonumber
\\
&+& \xi_2^t\Big(-2\Gamma^x_{mn}\nabla^m\xi_1^n + \nabla_m\nabla^m\xi_1^x
 - \nabla_m\nabla^x\xi_1^m \Big)
\nonumber
\\
&-& \xi_2^x\Big(-2\Gamma^t_{mn}\nabla^m\xi_1^n + \nabla_m\nabla^m\xi_1^t
 - \nabla_m\nabla^t\xi_1^m \Big)~.
\label{1.59}
\end{eqnarray}
Now since the integration (\ref{1.57}) will ultimately be evaluated on the horizon, we shall find the value of each term of the above very near the horizon. Therefore, using (\ref{1.06}), (\ref{1.51}) and the form of the generators (\ref{1.17}) we obtain the values of each term of the above expression near the horizon $x=0$ as,
\begin{eqnarray}
&&\nabla_m\xi_1^m \xi_2^t S^x = T_2\partial_t^2 T_1 - \frac{1}{2\kappa}\partial^2_t T_1\partial_t T_2 
\nonumber
\\
&&\xi_1^m\nabla_m\xi_2^t S^x = -\kappa T_1\partial_t T_2 + \kappa T_2 \partial_t T_1 + T_1 \partial^2_t T_2 - \frac{1}{2\kappa} \partial_t T_1\partial^2_t T_2
\nonumber
\\
&&- \xi_2^m\nabla_m\xi_1^tS^x = \kappa T_2\partial_t T_1 - \kappa T_1 \partial_t T_2 - T_2 \partial^2_t T_1 + \frac{1}{2\kappa} \partial_t T_2\partial^2_t T_1
\nonumber
\\
&&  -2\xi_2^t\Gamma^x_{mn}\nabla^m\xi_1^n = -T_2\partial^2_t T_1 + \frac{1}{2\kappa} \partial^2 T_1\partial_t T_2 
\nonumber
\\
&&\xi_2^t \nabla_m\nabla^m\xi_1^x = \frac{1}{2\kappa} T_2\partial^3_tT_1 - \frac{1}{4\kappa^2}\partial^3_tT_1\partial_tT_2 - 2\kappa T_2\partial_t T_1 + \partial_t T_1\partial_tT_2 + T_2\partial^2_t T_1 - \frac{1}{2\kappa}\partial^2_tT_1\partial_tT_2 
\nonumber
\\
&&- \xi_2^t\nabla_m\nabla^x\xi_1^m = 0 
\nonumber
\\
&& 2\xi_2^x\Gamma^t_{mn}\nabla^m\xi_1^n = -\frac{1}{2\kappa}\partial^2_t T_1\partial_t T_2 - 2\kappa x\partial_x T_1 \partial_t T_2 
\nonumber
\\
&& -\xi_2^x \nabla_m\nabla^m\xi_1^t = \frac{1}{4\kappa^2}\partial^3_t T_1\partial_tT_2 
\nonumber
\\
&& \xi_2^x\nabla_m\nabla^t\xi_1^m = - \frac{1}{4\kappa^2}\partial^3_t T_1\partial_tT_2 
\label{1.60}
\end{eqnarray}
So near the horizon (\ref{1.59}) reduces to
\begin{eqnarray}
K^{tx}_{12} &=& - 2\kappa T_1\partial_t T_2 + T_1\partial_t^2T_2 - \frac{1}{2\kappa} \Big(\partial_t T_1\partial^2_tT_2 + \partial^2_t T_1\partial_tT_2\Big)  +  \frac{1}{2\kappa} T_2\partial^3_tT_1 +  \partial_t T_1\partial_tT_2
\nonumber
\\
&& - \frac{1}{4\kappa^2}\partial^3T_1\partial_tT_2
\label{App29}
\end{eqnarray} 
Substituting this in (\ref{1.57}) and inserting the normalization factor, we obtain the expression for the bracket
\begin{eqnarray}
[Q[\xi_1],Q[\xi_2]]:&=& \frac{1}{16\pi G}\int_{\cal{H}}d^2x \Big[2\kappa\Big(T_1\partial_tT_2 - T_2\partial_tT_1\Big) - \Big(T_1\partial^2_tT_2 - T_2\partial^2_tT_1\Big)
\nonumber
\\
& +& \frac{1}{2\kappa}\Big(T_1\partial^3_tT_2 - T_2\partial^3_tT_1\Big) + \frac{1}{4\kappa^2}\Big(\partial^3_tT_1\partial_tT_2 - \partial^3_tT_2\partial_tT_1\Big)\Big]~.
\label{App30}
\end{eqnarray}
Similarly, (\ref{1.47}) yields,
\begin{eqnarray}
Q[\xi] = \frac{1}{8\pi G} \int d^2x \Big(\kappa T - \frac{1}{2}\partial_t T\Big)~.
\label{1.52}
\end{eqnarray}

  A couple of comments are in order. It must be noted that in finding the expression for the bracket (\ref{App30}), no use of boundary conditions (Dirichlet or Neumann) has been used. Earlier this was used for the case of $L_{bulk}$ to throw away the non-covariant terms in the bracket without giving any physical meaning \cite{Silva:2002jq}. Also, we did not use the condition $\delta_{\xi_1}\xi^a_2 =0$ (see Eq. (\ref{1.53})) which was adopted in earlier works. For instance, see \cite{Carlip:1999cy,Silva:2002jq}. This is logically correct since $\delta_{\xi_1}\xi^a_2 =0$ contradicts the algebra among the Fourier modes of the diffeomorphisms (see Eq. (\ref{4.11}), in next section).

\section{Virasoro algebra and entropy}
    In this section, the Fourier modes of the bracket and the charge will be found out. We shall show that for a particular ansatz for the the Fourier modes of the generators will lead to the Virasoro algebra. Finally using the Cardy formula, the entropy will be calculated.

    Consider the Fourier decompositions of $T_1$ and $T_2$:
\begin{eqnarray}
T_1 = \displaystyle\sum_{m} A_m T_m; \,\,\ T_2 = \displaystyle\sum_{n} B_n T_n~,
\label{4.10}
\end{eqnarray}  
where $A^*_m = A_{-m}$; $B^*_n = B_{-n}$. The Fourier modes $T_m$ will be chosen such that the Fourier modes of the diffeomorphisms (\ref{1.13}) obey one sub-algebra isomorphic to Diff. $S^1$:
\begin{eqnarray}
i\{\xi_m,\xi_n\}^a = (m-n)\xi_{m+n}^a~,
\label{4.11}
\end{eqnarray}
where $\{,\}$ is the Lie bracket. Now with the use of (\ref{4.10}), let us first find the Fourier modes of the bracket (\ref{App30}) and the charge (\ref{1.52}). Substitution of (\ref{4.10}) in (\ref{App30}) yields,
\begin{eqnarray}
[Q[\xi_1],Q[\xi_2]]:&=& \displaystyle\sum_{m,n}\frac{C_{m,n}}{16\pi G}\int_{\cal{H}}d^2x \Big[2\kappa\Big(T_m\partial_tT_n - T_n\partial_tT_m\Big) - \Big(T_m\partial^2_tT_n - T_n\partial^2_tT_m\Big)
\nonumber
\\
& +& \frac{1}{2\kappa}\Big(T_m\partial^3_tT_n - T_n\partial^3_tT_m\Big) + \frac{1}{4\kappa^2}\Big(\partial^3_tT_m\partial_tT_n - \partial^3_tT_n\partial_tT_m\Big)\Big]~,
\label{4.12}
\end{eqnarray}
where $C_{m,n} = A_m B_n$ and so $C^*_{m,n} = C_{-m,-n}$. Next defining the Fourier modes of $[Q[\xi_1],Q[\xi_2]]$ as
\begin{eqnarray}
[Q[\xi_1],Q[\xi_2]] = \displaystyle\sum_{m,n}C_{m,n}[Q_m,Q_n]~.
\label{4.13}
\end{eqnarray}
we find
\begin{eqnarray}
[Q_m,Q_n]: &=& \frac{1}{16\pi G}\int_{\cal{H}}d^2x \Big[2\kappa\Big(T_m\partial_tT_n - T_n\partial_tT_m\Big) - \Big(T_m\partial^2_tT_n - T_n\partial^2_tT_m\Big)
\nonumber
\\
& +& \frac{1}{2\kappa}\Big(T_m\partial^3_tT_n - T_n\partial^3_tT_m\Big) + \frac{1}{4\kappa^2}\Big(\partial^3_tT_m\partial_tT_n - \partial^3_tT_n\partial_tT_m\Big)\Big]~.
\label{4.14}
\end{eqnarray}
Similarly from (\ref{1.52}), the Fourier modes of the charge are given by
\begin{eqnarray}
Q_m = \frac{1}{8\pi G} \int d^2x \Big(\kappa T_m - \frac{1}{2}\partial_t T_m\Big)~,
\label{4.15}
\end{eqnarray}
where $Q[\xi] = \displaystyle\sum_m A_m Q_m$.
It must be noted that the present expression (\ref{4.15}) is exactly identical to that obtained in \cite{Majhi:2012tf} for the York-Gobbons-Hawking surface term whereas the other expression (\ref{4.14}) is different by some terms. This may be because these two surface terms are not exactly same. But we shall show that final result for the bracket is identical to the earlier analysis.  

  To calculate the above expressions (\ref{4.14}) and (\ref{4.15}) explicitly we need to have $T_m$'s. Following the earlier arguments, we choose
\begin{eqnarray}
T_m = \frac{1}{\alpha} e^{im\Big(\alpha t + g(x) + p.x_{\perp}\Big)}
\label{4.16}
\end{eqnarray}
such that they satisfy the algebra (\ref{4.11}). Here $\alpha$ is a constant, $p$ is an integer and $g(x) = G(X=x) = -\alpha \int\frac{dx}{2\kappa x}$. This is a standard choice in these computations and has been used several times in the literature \cite{Carlip:1998wz,Carlip:1999cy,Silva:2002jq}. It must be noted that the transverse directions are non-compact due to our Rindler approximations and so we will assume that $T_m$ is periodic in the transverse coordinates with the periodicities $L_y$ and $L_z$ on $y$ and $z$, respectively. Now substituting (\ref{4.16}) in (\ref{4.14}) and (\ref{4.15}) and then integrating over the cross-sectional area $A_{\perp}=L_yL_z$ we obtain
\begin{eqnarray}
&& Q_m = \frac{A_{\perp}}{8\pi G}\frac{\kappa}{\alpha}\delta_{m,0};
\label{4.17}
\\
&& i[Q_m,Q_n]: = \frac{A_{\perp}}{8\pi G}\frac{\kappa}{\alpha} (m-n)\delta_{m+n,0} + n^3 \frac{A_{\perp}}{16\pi G}\frac{\alpha}{\kappa}\delta_{m+n,0}~.
\label{4.18}
\end{eqnarray}
Using (\ref{4.17}), (\ref{4.18}) can be re-expressed as
\begin{eqnarray}
i[Q_m,Q_n]: = (m-n) Q_{m+n} + n^3 \frac{A_{\perp}}{16\pi G}\frac{\alpha}{\kappa}\delta_{m+n,0}~.
\label{4.19}
\end{eqnarray}
This is exactly identical to Virasoro algebra with the central charge $C$ is identified as
\begin{eqnarray}
\frac{C}{12} = \frac{A_{\perp}}{16\pi G}\frac{\alpha}{\kappa}~.
\label{4.20}
\end{eqnarray}
The zero mode eigenvalue is evaluated from (\ref{4.17}) for $m=0$:
\begin{eqnarray}
Q_0 = \frac{A_{\perp}}{8\pi G}\frac{\kappa}{\alpha}~.
\label{4.21}
\end{eqnarray}
Finally using the Cardy formula \cite{Cardy:1986ie,Carlip:1998qw}, we obtain the entropy as
\begin{eqnarray}
S = 2\pi\sqrt{\frac{CQ_0}{6}} = \frac{A_{\perp}}{4G}~,
\label{4.22}
\end{eqnarray}  
which is exactly the Bekenstein-Hawking entropy.

\section{Conclusions}
   It has already been observed that several interesting features and informations can be obtained from the surface term without incorporating the bulk term of the gravity action. In this paper we studied the surface term of the Einstein-Hilbert (EH) action in the context of Noether current. So far we know, this has not been attempted before. First the current was derived for an arbitrary diffeomorphism by using Noether prescription. Then we showed that the charge evaluated on the horizon for a Killing vector led to the Bekenstein-Hawking entropy after multiplied it by $2\pi/\kappa$. But till now, it is not known about the degrees of freedom responsible for the entropy. Here we addressed the issue and try to shed some light. This has been discussed in the context of Virasoro algebra and Cardy formula.

   In this paper, we defined the bracket among the charges. It was done, in the sprite of our earlier works \cite{Majhi:2011ws,Majhi:2012tf}, by taking the variation of the Noether potential $J^{ab}[\xi_1]$ for a different diffeomorphism $x^a\rightarrow x^a+\xi^a_2$, then an anti-symmetric combination between the indices $1$ and $2$ and integrating over the horizon surface. To achieve the final form, we did not use Einstein equation of motion or any ambiguous prescription, like vanishing of the variation of diffeomorphism parameter $\xi^a$, certain boundary conditions (e.g. Dirichlet or Neumann), etc. For explicit evaluation of our bracket, the spacetime was considered as the Rindler metric. The relevant diffeomorphisms were identified by using a very simple, physically motivated condition: {\it the diffeomorphisms keep the horizon structure of the metric invariant in some non-singular coordinate system}. It turned out that the Fourier modes of the bracket is similar to the Virasoro algebra. Identifying the central charge and the zero mode eigenvalue and then using these in Cardy formula we obtained exactly the Bekenstein-Hawking entropy.

   Let us now discuss in details what we have achieved in this paper. We first tabulate couple of technical points.
\vskip 1mm
\noindent
$\bullet$ To obtain the exact expression for entropy we did not need any hand waving prescriptions, like shifting of the value of the zero mode eigenvalue or the specific choice of the value of the parameter $\alpha$ appeared in Fourier modes of $T$ or both.
\vskip 1mm
\noindent
$\bullet$ The relevant diffeomorphisms for invariance of the horizon structure can be obtained various ways. Here our idea was to impose minimum constraints so that the bracket led to Virasoro algebra. It is also possible to have other choices of constraints to find the vectors $\xi^a$. For instance, the whole metric is invariant and the diffeomorphisms come out to be the Killing vectors which in general do not exist for a general spacetime.
\vskip 1mm
\noindent
Finally, we discuss several conceptual aspects. The analysis presents a nice connection between the horizon entropy and the degrees of freedom which are responsible for it. In the usual cases, one always find that the concepts of degrees of freedom and entropy are absolute. These does not have any observer dependent description. But in the case of gravity, as we know, the notion of temperature and entropy is observer dependent and hence one can expect that the degrees of freedom may  not be absolute. Here we showed that a certain class of observers which can see the horizon and keep the horizon structure invariant, always attribute entropy. This signifies the fact that among all the diffeomorphisms, some of them upgraded to real degrees of freedom which were originally gauge degrees of freedom and they have observer dependent notion. Also, everything what we achieved here, was done from surface term. This again illustrates the holographic nature of the gravity actions - {\it either the bulk and the surface terms may duplicate the same information or the surface term alone contains all the information about the theory of gravity}. Moreover, the methodology is general enough to discuss other theories of gravity \cite{Majhi}.   

\vskip 3mm
\noindent
{\bf Acknowledgement}\\
I thank T. Padmanabhan for several useful discussions.

\end{document}